\documentclass[prb,twocolumn,superscriptaddress,floatfix,amsmath,amssymb]{revtex4}
\usepackage{multirow}
\usepackage{graphicx}
\usepackage{bm}
\usepackage{dcolumn}
\usepackage{picture}
\usepackage[hidelinks]{hyperref}
\usepackage{natbib}
\usepackage{amsmath}
\usepackage{times}
\usepackage{color}
\usepackage{xcolor}
\usepackage{soul}
\usepackage{amsmath}
\usepackage{amssymb}

\begin{document}
	\title{Insulator-to-metal transition in the pyrochlore iridates series (Eu$ _{1-x}$Bi$_x$)$_2$Ir$ _2 $O$ _7$~probed using Hard X-ray Photoemission Spectroscopy }
	
	\author{Prachi Telang}
	\email[email:]{prachi.telang@students.iiserpune.ac.in}
	\affiliation{Department of Physics, Indian Institute of Science Education and Research, Pune, Maharashtra-411008, India}
	
	\author{Kshiti Mishra}
	\affiliation{Department of Physics, Indian Institute of Science Education and Research, Pune, Maharashtra-411008, India}
	
	\author{Rabindranath Bag}
	\affiliation{Department of Physics, Indian Institute of Science Education and Research, Pune, Maharashtra-411008, India}
	
	\author{A. Gloskovskii}
	\affiliation{Photon Science, Deutsches Elektronen-Synchrotron DESY, Notkestr. 85, 22607 Hamburg, Germany}
	
	\author{Yu. Matveyev}
	\affiliation{Photon Science, Deutsches Elektronen-Synchrotron DESY, Notkestr. 85, 22607 Hamburg, Germany}
	
	\author{Surjeet Singh}
	\email[email:]{surjeet.singh@iiserpune.ac.in}
	\affiliation{Department of Physics, Indian Institute of Science Education and Research, Pune, Maharashtra-411008, India}\affiliation{Center for Energy Science, Indian Institute of Science Education and Research, Pune, Maharashtra-411008, India}

\date{\today}

\begin{abstract}
Eu$ _2$Ir$ _2$O$ _7$, a candidate Weyl semimetal, shows an insulator-to-metal transition as a function of Bi substitution at the Eu site. In this work, we investigate the (Eu$ _{1-x}$Bi$_x$)$_2$Ir$ _2$O$ _7$ series via Hard X-ray Photoelectron Spectroscopy (HAXPES), where substitution of larger Bi$^{3+}$ for Eu$^{3+}$ is reported to result in an anomalous lattice contraction for $x \leqslant 0.035$. Using HAXPES we confirm that all the cations retain their nominal valence state throughout the series. The asymmetric nature of Bi core-level spectra for compositions in the metallic region indicates that Bi contributes to the density of states at the Fermi energy in this doping range. The valence band spectra show that the Bi $6s$ peak is unaltered throughout the series and is situated deep within the valence band. Instead, we argue that Bi $6p$--Ir $5d$ hybridisation drives the insulator-to-metal transition.
\end{abstract}

\pacs{}

\maketitle

\section{Introduction}
\label{Intro}
In pyrochlore oxides, A$_2$B$_2$O$_7$, with a $5d$ transition metal ion (e.g., Ir$ ^{4+} $) at the \textit{B}-site, the competing strengths of relativistic spin-orbit (SO) interaction, on-site Coulomb repulsion and crystal field splitting gives rise to non-trivial topological phases that have attracted considerable attention in the recent years \cite{WanWeyl, Krempa2014, Wang2017}. Due to SO interaction, the $t_{2g} $ level of Ir$^{4+}(d^5)$ in an octahedral crystal field splits into a completely-filled quadruplet (J$ _\textit{eff} = 3/2 $), and a higher-lying half-filled doublet (J$ _\textit{eff} = 1/2 $). Thus, an effective J$ _\textit{eff} = 1/2$ resides on the frustrated pyrochlore lattice, which is one of the key ingredients for realizing the interesting physical properties predicted for these compounds. The pyrochlores iridates exhibit a change of ground state from an antiferromagnetic(AFM)/insulator for smaller and heavier rare-earths (i.e., A$ =$ Gd, Tb, Dy, Ho, Er and Yb, including Y) to a correlated metal for the end member Pr$_2$Ir$_2$O$_7$, which remains magnetically unordered down to the lowest measurement temperature \cite{Tokiwa}. The intermediate members with A $=$ Nd, Sm and Eu, however, show a sharp metal-insulator transition (MI) concomitant with the onset of AFM long-range ordering of Ir$ ^{4+} $ moments \cite{Matsuhira2011}. The AFM state is of all-in/all-out type where all the Ir moments point into or away from the center of the tetrahedron formed by the nearest neighbor Ir atoms. This strong dependence of the ground state on the A-site ionic size is not so well understood. Also, while a variety of experiments have pointed to indirect signatures of exotic electronic phases in pyrochlore iridates (see for example Ref. [\citenum{Tafti, Sushkov, Ueda1}]), there exists very little direct experimental evidence. This underscores the importance of performing a detailed electronic structure study to throw light on the dependence of ground state properties on the A-site ionic radius or the lattice constant vis-\`{a}-vis corresponding changes in the electronic structure.

From among these pyrochlores, Eu$_2$Ir$_2$O$_7$ is of particular interest as it is the only member in this series with a non-magnetic A-site and yet located in the close proximity of the MI phase boundary. The non-magnetic nature of A-site is a consequence of the fact that in an Eu$ ^{3+}$(f$^7$) ion the spin ($S$) and orbital angular momenta ($L$) compensate each other, leading to a $J = 0$ ground state. Recently, some of us reported the magnetic and transport properties of (Eu$ _{1-x}$Bi$_x$)$_2$Ir$ _2$O$ _7$ series~\cite{Telang2019}. Similar to Eu$_2$Ir$_2$O$_7$, Bi$_2$Ir$_2$O$_7$ also crystallize with the pyrochlore structure with a nonmagnetic but larger-in-size Bi$^{3+}$ ion occupying the A-site. We found that contrary to the normal expectations, substitution of the larger Bi ion for Eu results in an anomalous lattice contraction for $x \leqslant 0.035$ without any change in the lattice symmetry. In this region, the ground state remains insulating and the MI transition, as well as the coincident AIAO ordering, becomes even more pronounced. In the range $0.05 \leqslant x \leqslant 0.1$, the MI or AIAO transition is rapidly suppressed, and for $x \geqslant 0.1$ a metallic ground state, akin to Pr$_2$Ir$_2$O$_7$, persists \cite{Telang2019}. While the transition from insulator to metal with Bi doping is not unexpected, the manner in which it occurs is rather interesting. In particular, we were intrigued by the observation of a large ($\simeq 20 \%$) negative lattice expansion, against which the ground state of Eu$_2$Ir$_2$O$_7$~remains robust. This motivated us to investigate and understand what drives the insulator-to-metal transition in this series, and what is the underlying reason for the anomalous lattice contraction. In particular, one of the questions we are asking is: Could the observed behavior be driven by a valence change of the Eu or Bi ions? Here, we study the oxidation state of cations and valence band spectra for different compositions using HAXPES, which is a powerful technique with a significant probing depth to overcome the surface effects, and hence the spectra obtained are representative of the bulk material. We establish that the anomalous lattice contraction in the range $0 \leqslant x \leqslant 0.035$ is not caused by a change in the oxidation state of either Bi or Eu. From the asymmetric shape of core-level spectra for Bi and Ir and restructuring of the valence band spectra in the metallic region, we argue that Bi $6p$--Ir $5d$ hybridisation drives the insulator-to-metal transition in this series.

\section{Experimental Details} 
\label{ExpDet}
All the samples in the series (Eu$_{1-x}$Bi$_x$)$_2$Ir$ _2$O$ _7$ were synthesized in air using solid-state synthesis method similar to the previous report\cite{Telang2019}. The phase purity of all the studied samples and their structural characterization was done using high-quality synchrotron data \cite{Telang2019}. As shown in Ref. \citenum{Telang2019}, the low-temperature resistivity of samples ($0 \leqslant x \leqslant 0.035$) in the insulting region is of the order of $10^3$ m$\Omega$ cm; it decreases to about $10$ m$\Omega$ cm ($0.05 \leqslant x \leqslant 0.1$) in the cross-over region, and finally to $1$ m$\Omega$ cm in the insulating region. HAXPES measurements were carried out at P22 beamline\cite{XPS} of PETRA III (DESY) utilizing Si$(311)$ double crystal monochromator. The excitation energy was E $= 5$ keV and the beam size at the sample $40 \times 20~\mu$m$^2$. Spectra were acquired with Specs $225$ HV analyzer, the overall energy resolution was set to $0.28$ eV.

\section{Results \& discussion}
\subsection{Europium core-level spectra}

\begin{figure}[!]
	\centering
	\includegraphics[width=0.44\textwidth]{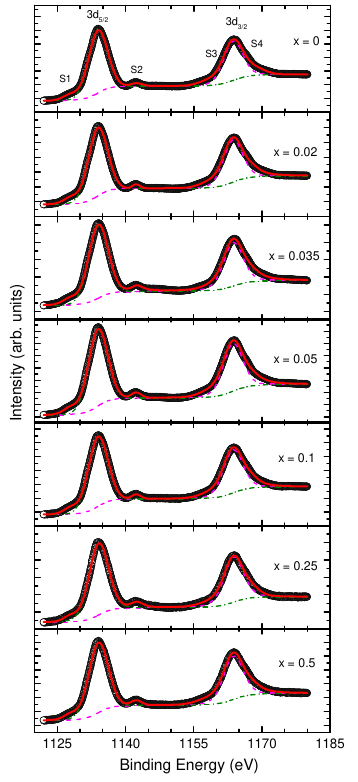}
	\caption{3d core-level spectra of Europium in series (Eu$ _{1-x}$Bi$_x$)$_2$Ir$ _2 $O$ _7$. The spectra consist of the standard 3d$_{3/2}$ and 3d$_{5/2}$ peaks along with three satellite peaks denoted by S1, S2, and S3. The shape and intensity of the peaks remain unchanged across the series implying lack of participation of Eu orbitals in driving the insulator-to-metal transition.}
	\label{1} 
\end{figure}

We start by presenting the core-level spectra for Europium. Similar to the rare-earth elements Ce, Pr, Sm, and Yb, Eu can also exhibit variable oxidation states. In the case of Eu, apart from the more commonly observed Eu$^{3+}$ (f$^6, J = 0$) oxidation state, Eu$^{2+}$ (f$^7, J = 7/2$) oxidation state is also known to exist in some compounds~\cite{Mercier}. Here, we probe the Eu $3$d core-level in the binding energy ranging from $1100$ to $1200$ eV. Fig. \ref{1} shows the $3d$ core-level spectra of Eu for various Bi doping concentrations. The precise values of binding energies for various peaks is given in Table \ref{T1}.\par

Due to the spin-orbit splitting, $3$d doublet corresponding to $3$d$_{3/2}$ and $3$d$_{5/2}$ is observed. These two peaks are separated in energy by $29.7$ eV, and are located at $\sim1163.6$ eV ($3$d$_{3/2}$) and $\sim1133.9$ eV ($3$d$_{5/2}$). Additionally, we also observe four small satellite peaks located at $1127.9$ eV (S1), $1142.4$ eV (S2), $1159$ eV (S3), and $1168.1$ eV (S4). From Fig. \ref{1}, one can see that the peak shape and peak positions remain unaltered with changing Bi doping concentration within the energy resolution of the detector. Also, the value of peak binding energies agrees well with that reported for Eu$^{3+}$~[\citenum{Caspers}]. This shows that Eu oxidation state as well as chemical environment is not affected by Bi substitution. The presence of satellites peaks can be attributed to the final state effect which arises due to the fact that in rare-earths, the empty 4f-subshells are lowered in energy through their interaction with the photoionized hole states. These 4f-subshells, thus, can trap the electrons, which leads to the appearance of ``shake-up" or ``shake-down" satellite peaks, depending on whether the electrons gain or lose energy, respectively. In europium compounds, the degeneracy of $4$f$^{6}$5d$^{1}$ and $4$f$^{7}$5d$^{0}$ configurations leads to the observation of both features in the final-state~\cite{Mercier}. Further, absence of any peak corresponding to divalent Eu in the d region, where the divalent and trivalent peaks are well separated (data not shown for brevity), further confirms that the additional peaks correspond to satellite peaks rather than divalent Eu.  Magnetic susceptibility data further supports this conclusion as for all compositions the measured susceptibility conforms to the trivalent nature of the Eu ion~\cite{Telang}.\par

\begin{table*}[!]
\setlength{\tabcolsep}{24 pt}
\caption{Eu 3d core-level peaks for various compositions ($x$) of the series (Eu$_{1-x}$Bi$_x$)$_2$Ir$ _2$O$ _7$. In addition to the core-level lines, there are four satellite peaks corresponding to S1, S2, S3, and S4.} 
\label{T1}
\begin{center}
	\resizebox{\textwidth}{!}{%
\begin{tabular}{c  c  c  c  c  c c}
  \hline
 \hspace*{0.1cm} $x$ \hspace*{0.1cm} & \hspace*{0.1cm} 3d$_{3/2}$ \hspace*{0.1cm} & \hspace*{0.1cm}3d$_{5/2}$ \hspace*{0.1cm} & \hspace*{0.1cm} S1 \hspace*{0.1cm} & \hspace*{0.1cm} S2 \hspace*{0.1cm} & \hspace*{0.1cm} S3 \hspace*{0.1cm}& \hspace*{0.1cm} S4 \hspace*{0.1cm}\\ 
\hline
\hline 
\hspace*{0.3cm}$0 $ & $ \hspace*{0.3cm}1163.6$ \hspace*{0.3cm} & $1133.9$ & $1127.9$ & $1142.4 $& $1159.0$ & $1168.1$\\
\hline 
\hspace*{0.3cm}$0.02 $ & $ \hspace*{0.3cm}1163.6$ \hspace*{0.3cm} & $1133.9$ & $1127.9$ & $1142.3$ & $1158.8$ &$1168.0$ \\
\hline
\hspace*{0.4cm}$0.035$ & \hspace*{0.3cm}$1163.7$ \hspace*{0.3cm} & $1134.0$ & $1127.8$ & $1142.5$ & $1159.0$ & $1168.1$\\ 
\hline 
\hspace*{0.2cm}$0.05$ & \hspace*{0.3cm}$1163.6$ \hspace*{0.3cm} & $1133.9$ & $1127.9$ & $1142.3$ & $1159.0$ & $1168.1$\\
\hline 
$0.1$ & \hspace*{0.3cm}$1163.6$ \hspace*{0.3cm} & $1133.9$ & $1127.9$ & $1142.4$ & $1158.8$ & $1167.9$\\
\hline 
\hspace*{0.2cm}$0.25$ & \hspace*{0.3cm}$1163.8$ \hspace*{0.3cm} & $1134.0$ & $1127.9$ & $1142.5$ & $1159.0$  & $1168.0$ \\
\hline 
\hspace*{0.2cm}$0.5$ & \hspace*{0.3cm}$1163.9$ \hspace*{0.3cm} & $1134.0$ & $1127.9$ & $1142.3$ & $1158.8$ & $1168.0$\\
\hline 
\end{tabular}
}
\end{center}
\end{table*}

\section{Iridium core-level spectra} 

The photoemission spectra of a number of Ir based compounds exhibit a distinct asymmetric shape, with satellite peaks associated with the core lines appearing in their neighborhood \cite{Kahk, Kennedy02, Freakley2017}. In previous XPS studies on pyrochlore iridates, these satellites were often attributed to the presence of higher oxidation states of Ir~\cite{Kennedy02}. However, in a recent study on Iridium oxide (IrO$_2$), Kahk et al.~\cite{Kahk}, reinvestigated this issue using HAXPES and density functional calculations. They show that the presence of peaks at higher binding energy do not originate from higher oxidation states. They ascribed these peaks to the final state effect, which we will discuss after showing our results and pointing out similarities and differences between our data and XPS or HAXPES data published previously.\par

\begin{figure}[!]
	\centering
	\includegraphics[width=0.44\textwidth]{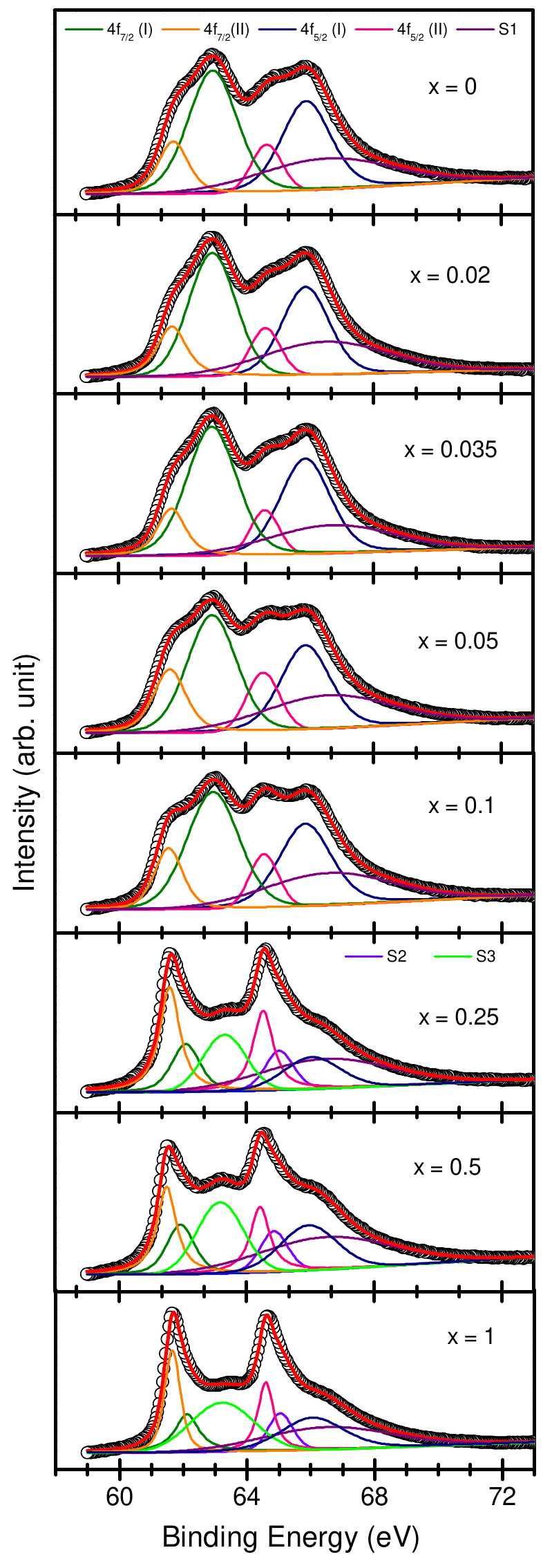}
	\caption{4f core-level photoemission spectra of Iridium for different compositions in the series (Eu$ _{1-x}$Bi$_x$)$_2$Ir$ _2 $O$ _7$. The spectrum shows significant changes in both shape and intensity of the peaks across the insulator-to-metal transition as discussed in the text.}
	\label{2}
\end{figure}

Fig.~\ref{2} shows Ir 4f core-level photoelectron spectra for various samples in the series. The spin-orbit coupling results in peaks corresponding to 4f$_{5/2}$ and 4f$_{7/2}$. However, each of these peaks is accompanied by a satellite peak at slightly lower binding energy. Altogether, we find that a complete deconvolution of the Ir photoelectron spectra requires five different peaks, which includes an additional satellite peak (S1) at relatively higher binding energy. Appropriate constraints corresponding to FWHM and area ratio were imposed while fitting the spin-orbit doublets. The obtained binding energies across the series for the twin Ir 4f$_{5/2}$ peaks fall in the energy range from $\approx 65.0$ eV to $\approx 65.9$ eV (4f$_{5/2}$~(I)) and from $\sim 64.4$ eV to $\approx 64.6$ eV (4f$_{5/2}$~(II)), whereas the Ir 4f$_{7/2}$ peaks ranged from $\approx 62.1$ eV to $\approx63$ eV (4f$_{7/2}$~(I)) and $61.5$ eV to $61.7$ eV (4f$_{7/2}$~(II)) for all the compositions. The additional satellite peak S1 is observed around $\approx66.6$ eV for all the compositions (see Table~\ref{T2} for details). 

In the insulating region ($0 \leq x \leq 0.035$) the spectra remains qualitatively unchanged, yielding a satisfactorily fit using the five peaks. Upon transiting into the crossover region ($0.05 \leq x \leq 0.01$), a slight increase in the intensity around $61.5$ eV and $64.5$ eV is observed but the spectra could still be fitted using the five peaks as before. In the core-level spectra for $x~\geqslant~0.1$ (i.e. for the highly metallic samples), the peak shape becomes increasingly asymmetric, mandating inclusion of two additional peaks to achieve a satisfactory fit. These peaks are labeled as S2 and S3 and they appear around $63$ eV and $66$ eV, respectively. Also, the components $4$f$_{5/2}$(II) and $4$f$_{7/2}$(II) of the spin-orbit doublet, that are broad and less intense up to $x = 0.035$, dominate the spectra for the metallic ($x$ $\geq$ $0.1$) samples. On the other hand, the components $4$f$_{5/2}$(I) and $4$f$_{7/2}$(I) dominant in the insulating region but are significantly suppressed in the metallic region. The HAXPES data of our Bi$_2$Ir$_2$O$_7$ agrees very well with that reported previously\cite{Sardar,Sun}, which rules out any extrinsic origin for these additional satellite peaks that are observed only for the metallic samples.\par

The changes upon going from insulating to metallic regime outlined above are analogous to those previously observed for some insulating and metallic ruthenate pyrochlores~\cite{Cox}. A qualitative understanding of these changes can be gained by considering that when a photoelectron leaves the core-level, the core-level gets ionized and creates an electrostatic perturbation at the ionized center. If the core-valence coulomb interaction exceeds the width of the one-electron conduction band, the ionized core will disengage one of the valence orbitals of the ionized atom from the conduction band~\cite{Campagna,Beatham}. The resulting localized atomic state lies below the Fermi energy, and can trap the emitted electron. Thus, photoelectrons with two different energies are detected, depending on whether the photoelectron was trapped (screened) or not (unscreened) in the localized energy state. This is also in agreement with a comprehensive DMFT study~\cite{Kim} on various ruthenates with varying degree of electronic correlations, where it was shown that the PE spectra exhibits a twin peak structure corresponding to the ``screened'' and ``unscreened'' components. Further, it was shown that the screened peak disappears in the Mott insulating state, but progressively develops as the band width increases and sample turns metallic. Similar changes in our Ir 4f core spectra are suggestive of a progressive conduction-band widening as the Bi doping concentration increases driving the system into a metallic state. To summarize this section, our experimental results find an excellent match with previous XPS studies on pyrochlore iridates~\cite{Pfeifer, Sun, Yang} and show that throughout the series Iridium retains its ${+4}$ oxidation state and the complex nature of the measured spectra arises, at least partly, due to ``screened" and ``unscreened" components rather than due to the presence of different oxidation states of Ir.


\begin{table*}[!]
\label{T2}
\centering
\caption{Ir 4f core-level peaks for all the measured compositions ($x$) in the series Eu$ _{1-x}$Bi$_x$)$_2$Ir$ _2 $O$ _7$. In addition to the core-level peaks corresponding to 4f$_{5/2}$(I \& II) and 4f$_{7/2}$(I \& II), there are additional satellite peaks named as S1, S2, and S3. The peaks S2 and S3 were required during the fitting of samples deep in the metallic regime i.e. $x \geq 0.25$ and can be ascribed to the final state effect (see text for details).} 
\begin{tabular*}{0.8\textwidth}{c @{\extracolsep{\fill}} c c c c c c c}
  \hline
$x$ & $4f_{5/2}$(I) & $4f_{5/2}$(II)  &  $4f_{7/2}$(I)  & $4f_{7/2}$(II) & S1 & S2 & S3 \\ 
\hline
\hline 
$0 $ & $65.88$  & $64.65$ & $62.95$ & $61.72$ & $66.53$ & - & - \\
\hline 
$0.02 $ & $65.87$ & $64.60$ & $62.94$ & $61.68$ & $66.50$ & - & - \\
\hline
$0.035$ & $65.86$ & $64.59$ & $62.93$ & $61.66$ & $66.71$ & - & - \\ 
\hline 
$0.05$ & $65.86$ & $64.54$ & $62.93$ & $61.60$ & $66.58$ & - & - \\
\hline 
$0.1$ & $65.84$ & $64.55$ & $62.95$ & $61.55$ & $66.57$ & - & - \\
\hline 
$0.25$ & $65.02$ & $64.52$ & $62.04$ & $61.59$ & $66.50$ & $66.05$ & $63.30$  \\
\hline 
$0.5$ & $64.86$ & $64.42$ & $61.93$ & $61.49$ & $66.50$ & $65.95$ & $63.17$\\
\hline 
$1$ & $65.04$ & $64.59$ & $62.11$ & $61.66$ & $66.49$ & $66.04$ &	$63.22$\\
\hline
\end{tabular*}

\end{table*}

\section{Bismuth core-level spectra}

\begin{figure}[!]
	\centering
	\includegraphics[width=0.44\textwidth]{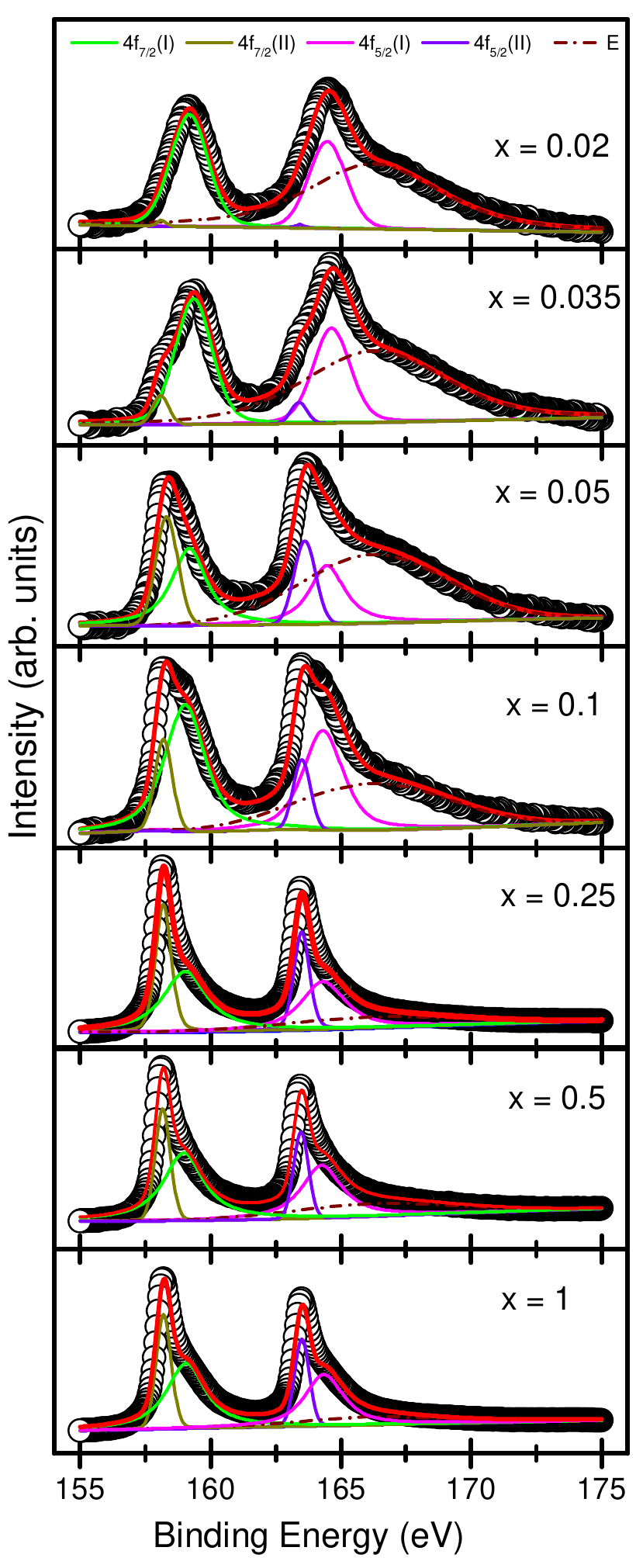}
	\caption{4f core-level spectra of Bismuth for various compositions ($x$) of the series (Eu$ _{1-x}$Bi$_x$)$_2$Ir$ _2 $O$ _7$. Changes in peak-shape and peak intensity with Bi doping is due to insulator-to-metal transition as discussed in the text.}
	\label{4}
\end{figure} 

Next, we probe the Bi 4f core-level to gain knowledge about its valence state. Bi can take valence states ranging from -3 to +5~\cite{Whitmire}. The spectra, plotted in Fig.~\ref{4}, show interesting changes with increasing Bi concentration. For all the compositions, the spectra could be deconvoluted using four peaks corresponding to two peaks each for the spin-orbit doublet 4f$_{5/2}$ and 4f$_{7/2}$, similar to the reports of other Bi based pyrochlore iridates where the twin peaks are not observed for insulating samples like Bi$_2$O$_3$, but are a typical feature of conducting samples where hybridization as discussed further plays a role~\cite{Cox,Sardar,Sun}. The peak position remains unchanged within the energy resolution throughout the series and matches well with previous reports for Bi$^{3+}$~\cite{Abdullah}. However, the relative peak intensity and peak shape exhibit significant changes upon going from insulating to metallic compositions. The peak positions for all the observed peaks are listed in Table~\ref{T3}.

\begin{table*}[!]
\label{T3}
\setlength{\tabcolsep}{28 pt}
\caption{Bi 4f core-level peaks for all the measured compositions ($x$) of the series Eu$_{2-2x}$Bi$_{2x}$Ir$_2$O$_7$. E represents the extraneous feature in the spectra which was held fixed at a given value while fitting the spectra for all the compositions.} 
\centering
\resizebox{\textwidth}{!}{%
\begin{tabular}{c  c  c  c  c  c }
  \hline
$x$ & $4f_{5/2}$(I) & $4f_{5/2}$(II)  &  $4f_{7/2}$(I)  & $4f_{7/2}$(II) & E \\ 
\hline
\hline 
$0.02 $ & $164.48$ & $163.42$ & $159.18$ & $158.12$ & $166.30$ \\
\hline
$0.035$ & $164.65$ & $163.39$ & $159.35$ & $158.09$ & $166.30$ \\ 
\hline 
$0.05$ & $164.47$ & $163.62$ & $159.17$ & $158.32$ & $166.30$ \\
\hline 
$0.1$ & $164.32$ & $163.50$ & $159.02$ & $158.20$ & $166.30$ \\
\hline 
$0.25$ & 164.32 & 163.5 & 159.02 & 158.2 & 166.3  \\
\hline 
$0.5$ & 164.26 & 163.47 & 158.96 & 158.17 & 166.3\\
\hline 
$1$ & 164.36 & 163.5 & 159.06 & 158.2 & 166.3\\
\hline
\end{tabular}
}
\end{table*}

For the insulating compositions ($x = 0.02$ and $0.035$), the peak shape is rather symmetric and the intensity of the components at lower binding energy (designated by 4f$_{5/2}$(I) and 4f$_{7/2}$(I)) is small compared to the higher binding energy components (designated by 4f$_{5/2}$(II) and 4f$_{7/2}$(II)). On the other hand, the peak shape becomes asymmetric and the intensity of 4f$_{5/2}$(I) and 4f$_{7/2}$(I) increases significantly upon entering the metallic regime, and at the same time the intensity of 4f$_{5/2}$(II) and 4f$_{7/2}$(II) decreases.\par

\begin{figure*}[!]
	\centering
	\includegraphics[width=0.9\textwidth]{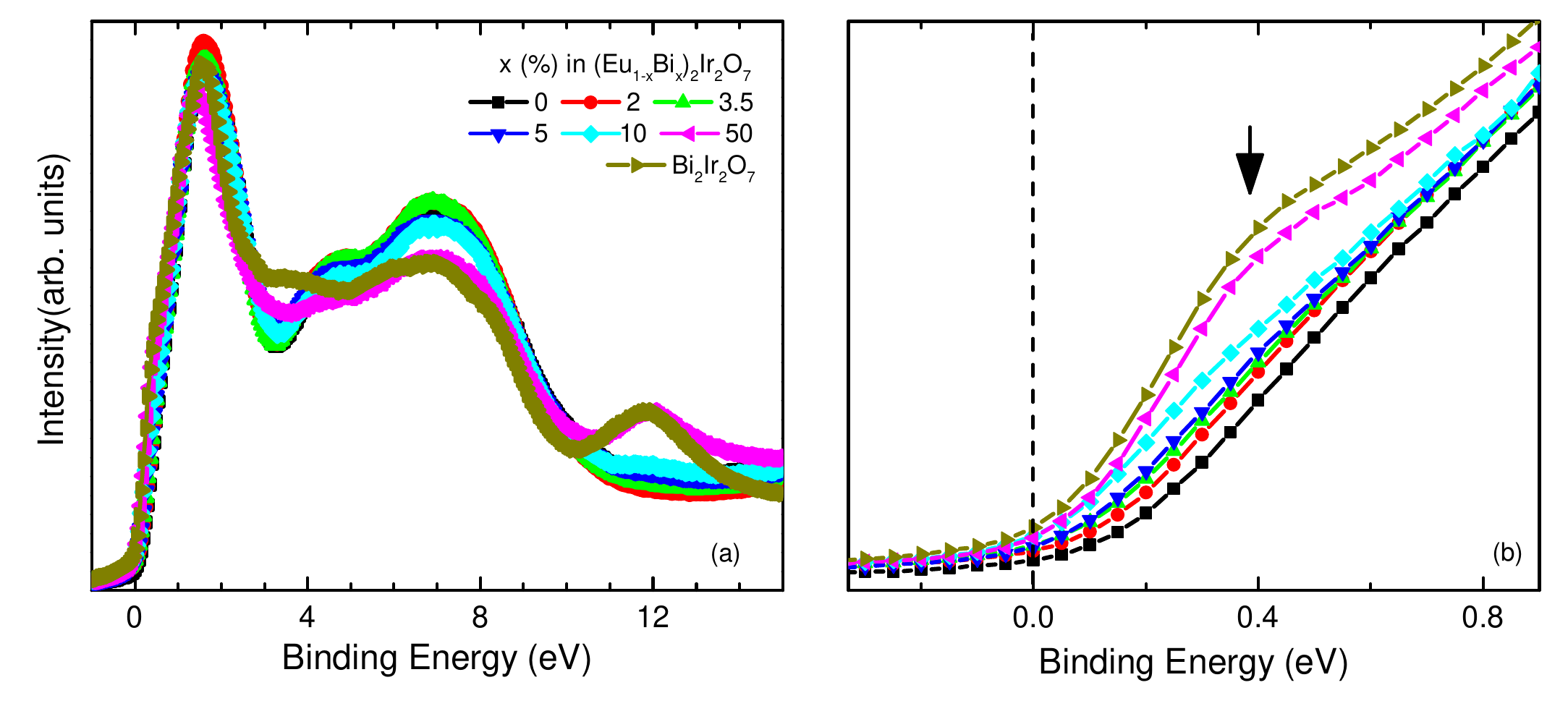}
	\caption{(a)Valence band spectra for different compositions of the series (Eu$_{1-x}$Bi$_x$)$_2$Ir$_2$O$_7$. The spectra shows significant changes in shape and intensity upon Bi substitution. (b) The region of the VBS in the vicinity of Fermi energy is shown on an enlarged scale.}
	\label{5}
\end{figure*} 

Apart from the standard Bi 4f doublets, we also observed an additional peak at around $166.3$ eV. As a peak at the same energy is observed even for Eu$_2$Ir$_2$O$_7$ (not shown here), which does not contain any Bi, we ascribe this feature to some extrinsic contribution. The peak position was kept fixed while fitting the spectra for all the compositions. Concerning the asymmetry of the core lines, the core photoelectron spectra of simple metals are known to display characteristic asymmetry due to electron-hole excitation in the final state~\cite{Cox, Wertheim}. Further, for metallic samples, it has been shown that the asymmetry of the core peak of a constituting element is proportional to the square of the partial density of states at the Fermi energy provided by the valence orbitals of that element~\cite{Folmer}. The increasing asymmetry of the core doublets in the Bi XPS spectra with insulator-to-metal transition suggests increasing contribution due to Bi at the Fermi energy.

\section{Valence Band Spectra}

We measured the valence band spectra (VBS) of all the samples to gain further insight into the insulator-to-metal transition with Bi substitution for Eu in Eu$_2$Ir$_2$O$_7$. VBS represents all the contributing bands near the Fermi level (E$_F$), which has been marked by $0$ eV in the spectra. The Fermi energy in our experiments was calibrated by measuring the VBS for a gold metal foil. In the VBS shown in Fig.~\ref{5}, normalized intensities are plotted. The normalization is done by dividing the measured intensity by the total integrated intensity for a given composition. The region from $0$ eV to $9$ eV is expected to be dominated by $5d$ orbitals of Ir~\cite{Sun}. On the other hand, the peak in VBS near $11$ eV, which is particularly becomes progressively pronounced as Bi-doping increases, can be attributed to Bi($6s^2$), as reported, for example, for Bi$_2$O$_3$ \cite{Walsh2006}. Since the $6s$ contribution is embedded deep below the Fermi energy, and its position remains invariant with Bi-doping, we can infer that Bi $6s$ hybridizes only weakly with Ir $5d$ orbitals. This suggestion is in line with previous photoemission and first principle study on Bi-based pyrochlore ruthenate Bi$_2$Ru$_2$O$_7$~\cite{Hsu}. Any significant contribution from O($2p$) in the measured VBS can also be ruled out due to its small photoionization cross-section at $5$ keV incident radiation. Now with $6s$ contribution due to Bi located deep within the valence band at 11 eV, the significant changes in the VBS near the Fermi energy on Bi-doping as highlighted in Fig.~\ref{5}b, can be attributed to possible hybridization between Ir($5d$) and Bi($6p$) orbitals in agreement with Ref. [\citenum{Hsu}]. A closer look near the Fermi energy (Fig.~\ref{5}b) reveals that the spectral weight gradually increases at the Fermi energy with increasing Bi doping, in agreement with the observation that with increasing Bi concentration the electrical conductivity increases i.e., the samples become more and more conducting. Also a shoulder peak appears near the Fermi energy, indicated with an arrow in Fig.~\ref{5}b, for $x \geq 0.1$, where the insulator-to-metal transition takes place in the series.\par


\section{Conclusion}

We present the core-level spectra of the cations and valence band spectra for all compositions in the series (Eu$_{1-x}$Bi$_x$)$_{2}$Ir$_2$O$_7$. We observed that the peak positions in the core-level spectra for all the cations only vary within the energy resolution of the spectrometer. This shows that all the cation retains their ideal valence state and importantly, the anomalous lattice contraction for $x \leq 0.035$ in the series (Eu$_{1-x}$Bi$_x$)$_{2}$Ir$_2$O$_7$ is not a result of varying oxidation state of any of the cations involved. However, the relative peak intensity and the peak shape vary considerably for the core-level spectra of both Ir and Bi. 

Going from insulating compositions for low Bi-doping to metallic compositions for high Bi doping, the core-level spectra for Ir and Bi shows pronounced asymmetry. The asymmetries of the Bi 4f and Ir 4f core signals provide direct evidence of an increasing partial density of states from these two elements at the Fermi energy. In addition to this, the appearance of Bi 6s peak without any shape or peak position change in the valence band spectra further supports our conjecture that metallicity in compositions above $x \gtrsim 0.1$ is driven by Bi $6p$--Ir$5d$ orbitals.\par

Our study elucidates the mechanism of metal-insulator transition in the series (Eu$_{1-x}$Bi$_x$)$_{2}$Ir$_2$O$_7$. This systemattic study tracks the changes in the core lines of the cations where isovalent substitution of Bi results in significant changes in physical and structural properties. Apart from ruling out the possibility of variable oxidation states, our study explains the debated role of Bi ($6s/6p$) hybridization in Bi-based metallic pyrochlore oxides. As Bi substitution preserves the Ir${4+}$ sublattice, it offers an avenue to realize non-trivial and possibly exotic metallic ground states in pyrochlore iridates and ruthenates.

\section*{Acknowledgments}
SS and PT are thankful to Prof. Sugata Ray for fruitful discussion. The authors acknowledges financial support from DST/SERB India under grant nos. EMR/2016/003792/PHY and SR/NM/TP-13/2016. PT and SS thank DST for financial support to perform experiments at Petra III, DESY. Funding for the HAXPES instrument at beamline P22 by the Federal Ministry of Education and Research (BMBF) under contracts 05KS7UM1 and 05K10UMA with Universit\"{a}t Mainz; 05KS7WW3, 05K10WW1 and 05K13WW1 with Universit\"{a}t W\"{u}rzburg is gratefully acknowledged. 
\nocite{}
\bibliography{Iridates_S}

\begin{thebibliography}{10}
\providecommand*{\bibinfo}[2]{#2}
\providecommand*{\eprint}[1]{#1}
\providecommand*{\url}[1]{#1}
\bibitem{WanWeyl}
\bibinfo{author}{X.~Wan}, \bibinfo{author}{A.~M. Turner},
  \bibinfo{author}{A.~Vishwanath}, and \bibinfo{author}{S.~Y. Savrasov},
  \bibinfo{journal}{Phys. Rev. B} \bibinfo{volume}{\textbf{83}},
  \bibinfo{pages}{205101} (\bibinfo{date}{May 2011}),
  \url{http://link.aps.org/doi/10.1103/PhysRevB.83.205101}.
\bibitem{Krempa2014}
\bibinfo{author}{W.~Witczak-Krempa}, \bibinfo{author}{G.~Chen},
  \bibinfo{author}{Y.~B. Kim}, and \bibinfo{author}{L.~Balents},
  \bibinfo{journal}{Annual Review of Condensed Matter Physics}
  \bibinfo{volume}{\textbf{5}}(1), \bibinfo{pages}{57} (\bibinfo{date}{2014}).
\bibitem{Wang2017}
\bibinfo{author}{R.~Wang}, \bibinfo{author}{A.~Go}, and \bibinfo{author}{A.~J.
  Millis}, \bibinfo{journal}{Phys. Rev. B} \bibinfo{volume}{\textbf{95}},
  \bibinfo{pages}{045133} (\bibinfo{date}{Jan 2017}).
\bibitem{Tokiwa}
\bibinfo{author}{Y.~Tokiwa}, \bibinfo{author}{J.~J. Ishikawa},
  \bibinfo{author}{S.~Nakatsuji}, and \bibinfo{author}{P.~Gegenwart},
  \bibinfo{journal}{Nature Materials} \bibinfo{volume}{\textbf{13}},
  \bibinfo{pages}{356} (\bibinfo{date}{2014}).
\bibitem{Matsuhira2011}
\bibinfo{author}{K.~Matsuhira}, \bibinfo{author}{M.~Wakeshima},
  \bibinfo{author}{Y.~Hinatsu}, and \bibinfo{author}{S.~Takagi},
  \bibinfo{journal}{Journal of the Physical Society of Japan}
  \bibinfo{volume}{\textbf{80}}(9), \bibinfo{pages}{094701}
  (\bibinfo{date}{2011}), \url{http://dx.doi.org/10.1143/JPSJ.80.094701}.
\bibitem{Tafti}
\bibinfo{author}{F.~F. Tafti}, \bibinfo{author}{J.~J. Ishikawa},
  \bibinfo{author}{A.~McCollam}, \bibinfo{author}{S.~Nakatsuji}, and
  \bibinfo{author}{S.~R. Julian}, \bibinfo{journal}{Phys. Rev. B}
  \bibinfo{volume}{\textbf{85}}, \bibinfo{pages}{205104} (\bibinfo{date}{May
  2012}).
\bibitem{Sushkov}
\bibinfo{author}{A.~B. Sushkov}, \bibinfo{author}{J.~B. Hofmann},
  \bibinfo{author}{G.~S. Jenkins}, \bibinfo{author}{J.~Ishikawa},
  \bibinfo{author}{S.~Nakatsuji}, \bibinfo{author}{S.~Das~Sarma}, and
  \bibinfo{author}{H.~D. Drew}, \bibinfo{journal}{Phys. Rev. B}
  \bibinfo{volume}{\textbf{92}}, \bibinfo{pages}{241108} (\bibinfo{date}{Dec
  2015}).
\bibitem{Ueda1}
\bibinfo{author}{K.~Ueda}, \bibinfo{author}{T.~Oh}, \bibinfo{author}{B.~J.
  Yang}, \bibinfo{author}{R.~Kaneko}, \bibinfo{author}{J.~Fujioka},
  \bibinfo{author}{N.~Nagaosa}, and \bibinfo{author}{Y.~Tokura},
  \bibinfo{journal}{Nature Communications} \bibinfo{volume}{\textbf{8}},
  \bibinfo{pages}{15515} (\bibinfo{date}{May 2017}).
\bibitem{Telang2019}
\bibinfo{author}{P.~Telang}, \bibinfo{author}{K.~Mishra},
  \bibinfo{author}{G.~Prando}, \bibinfo{author}{A.~K. Sood}, and
  \bibinfo{author}{S.~Singh}, \bibinfo{journal}{Phys. Rev. B}
  \bibinfo{volume}{\textbf{99}}, \bibinfo{pages}{201112} (\bibinfo{date}{May
  2019}), \url{https://link.aps.org/doi/10.1103/PhysRevB.99.201112}.
\bibitem{XPS}
\bibinfo{author}{C.~Schlueter}, \bibinfo{author}{A.~Gloskovskii},
  \bibinfo{author}{K.~Ederer}, \bibinfo{author}{I.~Schostak},
  \bibinfo{author}{S.~Piec}, \bibinfo{author}{I.~Sarkar},
  \bibinfo{author}{Y.~Matveyev}, \bibinfo{author}{P.~Lömker},
  \bibinfo{author}{M.~Sing}, \bibinfo{author}{R.~Claessen}, \emph{et~al.},
  \bibinfo{journal}{AIP Conference Proceedings}
  \bibinfo{volume}{\textbf{2054}}(1), \bibinfo{pages}{040010}
  (\bibinfo{date}{2019}),
  \url{https://aip.scitation.org/doi/abs/10.1063/1.5084611}.
\bibitem{Mercier}
\bibinfo{author}{F.~Mercier}, \bibinfo{author}{C.~Alliot},
  \bibinfo{author}{L.~Bion}, \bibinfo{author}{N.~Thromat}, and
  \bibinfo{author}{P.~Toulhoat}, \bibinfo{journal}{Journal of Electron
  Spectroscopy and Related Phenomena} \bibinfo{volume}{\textbf{150}}(1),
  \bibinfo{pages}{21 } (\bibinfo{date}{2006}),
  \url{http://www.sciencedirect.com/science/article/pii/S0368204805004366}.
\bibitem{Caspers}
\bibinfo{author}{C.~Caspers}, \bibinfo{author}{M.~M\"uller},
  \bibinfo{author}{A.~X. Gray}, \bibinfo{author}{A.~M. Kaiser},
  \bibinfo{author}{A.~Gloskovskii}, \bibinfo{author}{C.~S. Fadley},
  \bibinfo{author}{W.~Drube}, and \bibinfo{author}{C.~M. Schneider},
  \bibinfo{journal}{Phys. Rev. B} \bibinfo{volume}{\textbf{84}},
  \bibinfo{pages}{205217} (\bibinfo{date}{Nov 2011}).
\bibitem{Telang}
\bibinfo{author}{P.~Telang}, \bibinfo{author}{K.~Mishra},
  \bibinfo{author}{A.~K. Sood}, and \bibinfo{author}{S.~Singh},
  \bibinfo{journal}{Phys. Rev. B} \bibinfo{volume}{\textbf{97}},
  \bibinfo{pages}{235118} (\bibinfo{date}{Jun 2018}).
\bibitem{Kahk}
\bibinfo{author}{J.~M. Kahk}, \bibinfo{author}{C.~G. Poll},
  \bibinfo{author}{F.~E. Oropeza}, \bibinfo{author}{J.~M. Ablett},
  \bibinfo{author}{D.~C\'eolin}, \bibinfo{author}{J.-P. Rueff},
  \bibinfo{author}{S.~Agrestini}, \bibinfo{author}{Y.~Utsumi},
  \bibinfo{author}{K.~D. Tsuei}, \bibinfo{author}{Y.~F. Liao}, \emph{et~al.},
  \bibinfo{journal}{Phys. Rev. Lett.} \bibinfo{volume}{\textbf{112}},
  \bibinfo{pages}{117601} (\bibinfo{date}{Mar 2014}).
\bibitem{Kennedy02}
\bibinfo{author}{B.~J. Kennedy}, \bibinfo{journal}{Physica B: Condensed Matter}
  \bibinfo{volume}{\textbf{241}}, \bibinfo{pages}{303 } (\bibinfo{date}{1997}).
\bibitem{Freakley2017}
\bibinfo{author}{S.~J. Freakley}, \bibinfo{author}{J.~Ruiz-Esquius}, and
  \bibinfo{author}{D.~J. Morgan}, \bibinfo{journal}{Surface and Interface
  Analysis} \bibinfo{volume}{\textbf{49}}(8), \bibinfo{pages}{794}
  (\bibinfo{date}{2017}),
  \eprint{https://onlinelibrary.wiley.com/doi/pdf/10.1002/sia.6225},
  \url{https://onlinelibrary.wiley.com/doi/abs/10.1002/sia.6225}.
\bibitem{Sardar}
\bibinfo{author}{K.~Sardar}, \bibinfo{author}{S.~C. Ball},
  \bibinfo{author}{J.~D. Sharman}, \bibinfo{author}{D.~Thompsett},
  \bibinfo{author}{J.~M. Fisher}, \bibinfo{author}{R.~A. Smith},
  \bibinfo{author}{P.~K. Biswas}, \bibinfo{author}{M.~R. Lees},
  \bibinfo{author}{R.~J. Kashtiban}, \bibinfo{author}{J.~Sloan}, \emph{et~al.},
  \bibinfo{journal}{Chemistry of Materials} \bibinfo{volume}{\textbf{24}}(21),
  \bibinfo{pages}{4192} (\bibinfo{date}{2012}),
  \eprint{https://doi.org/10.1021/cm302468b},
  \url{https://doi.org/10.1021/cm302468b}.
\bibitem{Sun}
\bibinfo{author}{W.~Sun}, \bibinfo{author}{J.-Y. Liu}, \bibinfo{author}{X.-Q.
  Gong}, \bibinfo{author}{W.-Q. Zaman}, \bibinfo{author}{L.-M. Cao}, and
  \bibinfo{author}{J.~Yang}, \bibinfo{journal}{Scientific Reports}
  \bibinfo{volume}{\textbf{6}}, \bibinfo{pages}{38429} (\bibinfo{date}{Dec
  2016}).
\bibitem{Cox}
\bibinfo{author}{P.~A. Cox}, \bibinfo{author}{R.~G. Egdell},
  \bibinfo{author}{J.~B. Goodenough}, \bibinfo{author}{A.~Hamnett}, and
  \bibinfo{author}{C.~C. Naish}, \bibinfo{journal}{Journal of Physics C: Solid
  State Physics} \bibinfo{volume}{\textbf{16}}(32), \bibinfo{pages}{6221}
  (\bibinfo{date}{nov 1983}).
\bibitem{Campagna}
\bibinfo{author}{M.~Campagna}, \bibinfo{author}{G.~K. Wertheim},
  \bibinfo{author}{H.~R. Shanks}, \bibinfo{author}{F.~Zumsteg}, and
  \bibinfo{author}{E.~Banks}, \bibinfo{journal}{Phys. Rev. Lett.}
  \bibinfo{volume}{\textbf{34}}, \bibinfo{pages}{738} (\bibinfo{date}{Mar
  1975}), \url{https://link.aps.org/doi/10.1103/PhysRevLett.34.738}.
\bibitem{Beatham}
\bibinfo{author}{N.~Beatham}, \bibinfo{author}{P.~Cox},
  \bibinfo{author}{R.~Egdell}, and \bibinfo{author}{A.~Orchard},
  \bibinfo{journal}{Chemical Physics Letters} \bibinfo{volume}{\textbf{69}}(3),
  \bibinfo{pages}{479 } (\bibinfo{date}{1980}).
\bibitem{Kim}
\bibinfo{author}{B.~J. Kim}, \bibinfo{author}{H.~Jin}, \bibinfo{author}{S.~J.
  Moon}, \bibinfo{author}{J.-Y. Kim}, \bibinfo{author}{B.-G. Park},
  \bibinfo{author}{C.~S. Leem}, \bibinfo{author}{J.~Yu}, \bibinfo{author}{T.~W.
  Noh}, \bibinfo{author}{C.~Kim}, \bibinfo{author}{S.-J. Oh}, \emph{et~al.},
  \bibinfo{journal}{Phys. Rev. Lett.} \bibinfo{volume}{\textbf{101}},
  \bibinfo{pages}{076402} (\bibinfo{date}{Aug 2008}).
\bibitem{Pfeifer}
\bibinfo{author}{V.~Pfeifer}, \bibinfo{author}{T.~E. Jones},
  \bibinfo{author}{J.~J. Velasco~Velez}, \bibinfo{author}{C.~Massue},
  \bibinfo{author}{R.~Arrigo}, \bibinfo{author}{D.~Teschner},
  \bibinfo{author}{F.~Girgsdies}, \bibinfo{author}{M.~Scherzer},
  \bibinfo{author}{M.~T. Greiner}, \bibinfo{author}{J.~Allan}, \emph{et~al.},
  \bibinfo{journal}{Surface and Interface Analysis}
  \bibinfo{volume}{\textbf{48}}(5), \bibinfo{pages}{261}
  (\bibinfo{date}{2016}).
\bibitem{Yang}
\bibinfo{author}{W.~C. Yang}, \bibinfo{author}{Y.~T. Xie},
  \bibinfo{author}{W.~K. Zhu}, \bibinfo{author}{K.~Park},
  \bibinfo{author}{A.~P. Chen}, \bibinfo{author}{Y.~Losovyj},
  \bibinfo{author}{Z.~Li}, \bibinfo{author}{H.~M. Liu},
  \bibinfo{author}{M.~Starr}, \bibinfo{author}{J.~A. Acosta}, \emph{et~al.},
  \bibinfo{journal}{Scientific Reports} \bibinfo{volume}{\textbf{7740}}
  (\bibinfo{date}{Aug 2017}).
\bibitem{Whitmire}
\bibinfo{author}{K.~H. Whitmire}, \bibinfo{title}{\emph{Bismuth: Inorganic
  Chemistry}} (\bibinfo{publisher}{American Cancer Society},
  \bibinfo{year}{2014}), ISBN \bibinfo{isbn}{9781119951438}.
\bibitem{Abdullah}
\bibinfo{author}{E.~A. Abdullah} and \bibinfo{author}{T.~K. Ban},
  \bibinfo{journal}{E-Journal of Chemistry} \bibinfo{volume}{\textbf{9}},
  \bibinfo{pages}{2429} (\bibinfo{date}{Jan 2012}).
\bibitem{Wertheim}
\bibinfo{author}{P.~Day}, \bibinfo{title}{\emph{Emission and Scattering
  Techniques: Studies of Inorganic Molecules, Solids, and Surfaces}},
  \bibinfo{volume}{Nato Science Series C:} (\bibinfo{publisher}{Springer
  Netherlands}, \bibinfo{year}{2012}), ISBN \bibinfo{isbn}{9789400985254}.
\bibitem{Folmer}
\bibinfo{author}{J.~Folmer} and \bibinfo{author}{D.~de~Boer},
  \bibinfo{journal}{Solid State Communications}
  \bibinfo{volume}{\textbf{38}}(12), \bibinfo{pages}{1135 }
  (\bibinfo{date}{1981}).
\bibitem{Walsh2006}
\bibinfo{author}{A.~Walsh}, \bibinfo{author}{G.~W. Watson},
  \bibinfo{author}{D.~J. Payne}, \bibinfo{author}{R.~G. Edgell},
  \bibinfo{author}{J.~Guo}, \bibinfo{author}{P.-A. Glans},
  \bibinfo{author}{T.~Learmonth}, and \bibinfo{author}{K.~E. Smith},
  \bibinfo{journal}{Phys. Rev. B} \bibinfo{volume}{\textbf{73}},
  \bibinfo{pages}{235104} (\bibinfo{date}{Jun 2006}),
  \url{https://link.aps.org/doi/10.1103/PhysRevB.73.235104}.
\bibitem{Hsu}
\bibinfo{author}{W.~Y. Hsu}, \bibinfo{author}{R.~V. Kasowski},
  \bibinfo{author}{T.~Miller}, and \bibinfo{author}{T.~Chiang},
  \bibinfo{journal}{Applied Physics Letters} \bibinfo{volume}{\textbf{52}}(10),
  \bibinfo{pages}{792} (\bibinfo{date}{1988}).

\end{thebibliography}
\bibliographystyle{revtex}

\pagebreak
\end{document}